\documentclass[10pt]{article}  
\usepackage{epsf}
\usepackage{graphics}
\usepackage{psfig}
\catcode`\@=11

\@addtoreset{equation}{section}
\def\theequation{\arabic{section}.\arabic{equation}}
\catcode`\@=11

\def\section{\@startsection{section}{1}{\z@}{3.5ex plus 1ex minus
   .2ex}{2.3ex plus .2ex}{\large\bf}}

\def\eqnarray{\let\@currentlabel=\theequation\refstepcounter{equation}
    \global\@eqnswtrue
    \global\@eqcnt\z@\tabskip\@centering\let\\=\@eqncr
    $$\halign to \displaywidth\bgroup\@eqnsel\hskip\@centering
      $\displaystyle\tabskip\z@{##}$&\global\@eqcnt\@ne
       \hfil${{}##{}}$\hfil
      &\global\@eqcnt\tw@ $\displaystyle\tabskip\z@{##}$\hfil
       \tabskip\@centering&\llap{##}\tabskip\z@\cr}
\def\lefteqn#1{\hbox to 4\arraycolsep{$\displaystyle #1$\hss}}
\def\thesection{\arabic{section}.}

\def\appendix{\setcounter{section}{0}
        \def\thesection{Appendix.}
        \def\theequation{\Alph{section}.\arabic{equation}}}

\long\def\@makefntext#1{\parindent 0cm\noindent
\hbox to 1em{\hss$^{\@thefnmark}$}#1}
\def\IR{{\hbox{{\rm I}\kern-.2em\hbox{\rm R}}}}
\def\IH{{\hbox{{\rm I}\kern-.2em\hbox{\rm H}}}}
\def\IC{{\ \hbox{{\rm I}\kern-.6em\hbox{\bf C}}}}
\def\IZ{{\hbox{{\rm Z}\kern-.4em\hbox{\rm Z}}}}

\newcommand{\beq}{\begin{equation}}
\newcommand{\be}{\begin{equation}}
\newcommand{\eeq}{\end{equation}}
\newcommand{\ee}{\end{equation}}
\newcommand{\bea}{\begin{eqnarray}}
\newcommand{\eea}{\end{eqnarray}}
\newcommand{\bean}{\begin{eqnarray*}}
\newcommand{\eean}{\end{eqnarray*}}
\newcommand{\ba}{\beq\begin{array}{lll} }
\newcommand{\ea}{\end{array}\eeq}

\def\IC{ {\rm l\hspace{-1.2ex}C} }    %  mine is better....
\def\IZ{{\hbox{{\rm Z}\kern-.4em\hbox{\rm Z}}}} 
\def\IR{{\hbox{{\rm I}\kern-.2em\hbox{\rm R}}}} 
\begin{document}                           %
                                                                 %
                                                                 %
%%%%%%%%%%%%%%%%%%%%%%%%%%%%%%%%%%%%%%%%%
%     C I T E . S T Y
%     compressed lists of numerical citations: [11-16]
%     see also OVERCITE.STY and DRFTCITE.STY
%
%     Copyright (C) 1989-1992 by Donald Arseneau
%     These macros may be freely transmitted, reproduced, or modified for
%     non-commercial purposes provided that this notice is left intact.
%
%
%  \@citen contains the code that parses the list of names, ignoring
%  spaces after commas, writes the aux file \citation, and formats the
%  number list.  \citen can be used by itself to give citation numbers
%  without the other formatting; e.g., "See also ref.~\citen{junk}."
%
\def\citen#1{%
\edef\@tempa{\@ignspaftercomma,#1, \@end, }% ignore spaces in parameter list
\edef\@tempa{\expandafter\@ignendcommas\@tempa\@end}%
\if@filesw \immediate \write \@auxout {\string \citation {\@tempa}}\fi
\@tempcntb\m@ne \let\@h@ld\relax \let\@citea\@empty
\@for \@citeb:=\@tempa\do {\@cmpresscites}%
\@h@ld}
%
% for ignoring spaces in the input:
\def\@ignspaftercomma#1, {\ifx\@end#1\@empty\else
   #1,\expandafter\@ignspaftercomma\fi}
\def\@ignendcommas,#1,\@end{#1}
%
% For each citation, check if it is defined, if it is a number, and
% if it is a consecutive number    that can be represented like 3-7.
%
\def\@cmpresscites{%
 \expandafter\let \expandafter\@B@citeB \csname b@\@citeb \endcsname
 \ifx\@B@citeB\relax % undefined
    \@h@ld\@citea\@tempcntb\m@ne{\bf ?}%
    \@warning {Citation `\@citeb ' on page \thepage \space undefined}%
 \else%  defined
    \@tempcnta\@tempcntb \advance\@tempcnta\@ne
    \setbox\z@\hbox\bgroup % check if citation is a number:
    \ifnum\z@<0\@B@citeB \relax
       \egroup \@tempcntb\@B@citeB \relax
       \else \egroup \@tempcntb\m@ne \fi
    \ifnum\@tempcnta=\@tempcntb % Number follows previous--hold on to it
       \ifx\@h@ld\relax % first pair of successives
          \edef \@h@ld{\@citea\@B@citeB}%
       \else % compressible list of successives
%         % use \hbox to avoid easy \exhyphenpenalty breaks
          \edef\@h@ld{\hbox{--}\penalty\@highpenalty \@B@citeB}%
       \fi
    \else   %  non-successor--dump what's held and do this one
       \@h@ld \@citea \@B@citeB \let\@h@ld\relax
 \fi\fi%
 \let\@citea\@citepunct
}
%
%%    To put space after the comma, use:
\def\@citepunct{,\penalty\@highpenalty\hskip.13em plus.1em minus.1em}%
%%    For no space after comma, use:
%% \def\@citepunct{,\penalty\@highpenalty}%
%%
%
%  Make \@citex refer to \citen:
%
\def\@citex[#1]#2{\@cite{\citen{#2}}{#1}}%
%
%  Replacement for \@cite.  Give one normal space before the citation,
%  set high penalties for linebreaks,
%
\def\@cite#1#2{\leavevmode\unskip
  \ifnum\lastpenalty=\z@ \penalty\@highpenalty \fi % highpenalty before
  \ [{\multiply\@highpenalty 3 #1% % triple-highpenalties within list
      \if@tempswa,\penalty\@highpenalty\ #2\fi % and before note.
    }]\spacefactor\@m}
\let\nocitecount\relax  % in case \nocitecount was used for drftcite
%%%%%%%%%%%%%%%%%%%%%%%%%%%%%%%%%%%%%%%%%%%%%%%%%%%%%%%%%%%%%%%%%%%%%%%
%%%%%%%%%%%%%%%%%%%%%%%%%%%%%%%%%%%%%%%%%%%%%%%%%%%%%%%%%%%%%%%%%%%%%%%
%                          BODY                                       %
%%%%%%%%%%%%%%%%%%%%%%%%%%%%%%%%%%%%%%%%%%%%%%%%%%%%%%%%%%%%%%%%%%%%%%%
\title{GIST: A tool for Global Ionospheric Tomography using GPS ground and LEO data and sources of
opportunity with applications in instrument calibration}
\author{A. Flores, G. Ruffini, A. Rius, and E. Cardellach}
\maketitle

\begin{abstract}
Ionospheric tomography using GPS data has been reported in the literature  and
even the application to radar altimeter calibration was succesfully carried out in a recent work 
(\cite{199ruffini1998}). We here
present a new software tool, called Global Ionospheric Stochastic Tomography software (GIST), 
and its powerful capability for ingesting GPS data from different sources
(ground stations, receivers on board LEO for navigation and occultation purposes) and other data such as altimetry data
to yield global maps with dense coverage and inherent calibration of the instruments. We show results obtained 
including 106 IGS ground stations, GPS/MET low rate occultation data,
TOPEX/POSEIDON GPS data from the navigation antenna and NASA Radar Altimeter with the additional benefit
of a direct estimation of the NRA bias. The possibility of ingesting different kinds of ionospheric data 
into the tomographic model suggest a way to accurately monitor the ionosphere with direct application to 
single frequency instrument calibration.
\end{abstract}

\section{Introduction}

Radio waves traversing the ionosphere suffer a delay of a well-known
dispersive nature and it is common to suppress this effect by using a combination of signals at two
separated frequencies. However, there are two aspects here to be considered: first, the electronic 
equipment of on board instrumentation  has to be periodically
calibrated, and second, duplicating systems to operate at two frequencies adds cost and complexity to the instruments.
Therefore it is desirable to have a system able to reproduce the status of the ionosphere, and use it for monitoring,
and single- and dual-frequency instrument calibration.
Tomographic techniques are applied to this end ingesting data from different sources. In previous
references \cite{75rius1997}, \cite{156ruffini1998}, \cite{199ruffini1998} we have discussed the 
tomographic methodology and some different implementations, which we will here briefly summarize. 
This work intends to highlight the successful elaboration of 
a software package that implements those techniques and also to emphasize the possibility of ingesting data other than GPS
to densify the receivers network.

\section{Tomographic technique}
The ionospheric delay can be determined in a bistatic dual-frequency system from phase measurements following the equation:
\bea
L_I(\vec r, t) =L_1-L_2=\gamma\int_{ray}dl \rho(\vec r, t)+c_r+c_t,
\label{ionos_comb}
\eea
where we have noted the phase measurements with $L$. The factor $\gamma$ depends on the frequencies in use (for
GPS $\gamma=1.05\cdot 10^{-17}$ m$^3$/el) and $\rho$ is the electron density. The two constants $c_r$ and $c_t$
are the biases associated to the transmitter and receiver (\cite{79sardon1994}).
Tomographic analysis consist in obtainting the solution fields ($\rho$) from the integrated value along the ray paths and 
Equation \ref{ionos_comb} is termed as the ``tomographic equation''. If $\rho$ is expressed as a linear combination of a set
of basis functions $\rho=\sum_j x_j(t) \Psi_j(\vec r) + \epsilon(\vec r, t)$ then the above equation becomes 
$L_I=y_i=\sum_Jx_J(t)\int_{s.l.}\Psi_J(\vec r) d\vec l +\zeta(\vec r,t)+c_r+c_t$ and can be written for each 
ray to obtain a set of linear equations such as ${\bf y} = {\bf A}\cdot {\bf x}$. In our tomographic 
system, we choose voxels as the basis functions. Voxels are 3-D pixels or fuctions valued 1 inside the volume 
of the voxel and 0 elsewhere. Empirical Orthogonal Functions can also
be used as shown in \cite{203howe1998}.The system, however, may not have a solution because data are not
uniformly distributed, and thus we seek to
minimize the functional
\bea
\chi^2(x)=(y-Ax)^T\cdot (y-Ax).
\eea
In  \cite{156ruffini1998} we discussed the use of a correlation functional to confine the spatial spectrum of the
solution to the low portion of the frequency space. The same concept can be expressed by adding new equations 
({\it constraints}) that impose that the density in a voxel be a weighted average of its neighbours (\cite{75rius1997}).
To take into account variation in time, a Kalman filter is implemented, considering the density to behave as
a random walk stochastic process. Instrumental constants are also considered and resolved as constants
or eliminated by differencing \cite{79sardon1994}, \cite{213hernandez1998}. While differencing reduces the number of
unknowns, estimation furnishes the solution with more information and provides nuisance parameters to absorb
noise from the system.
\section{The GIST tool for ionospheric tomography}
\begin{figure}
\begin{center}

\mbox{
\rotatebox{90}{
\epsfxsize=15cm
\epsffile{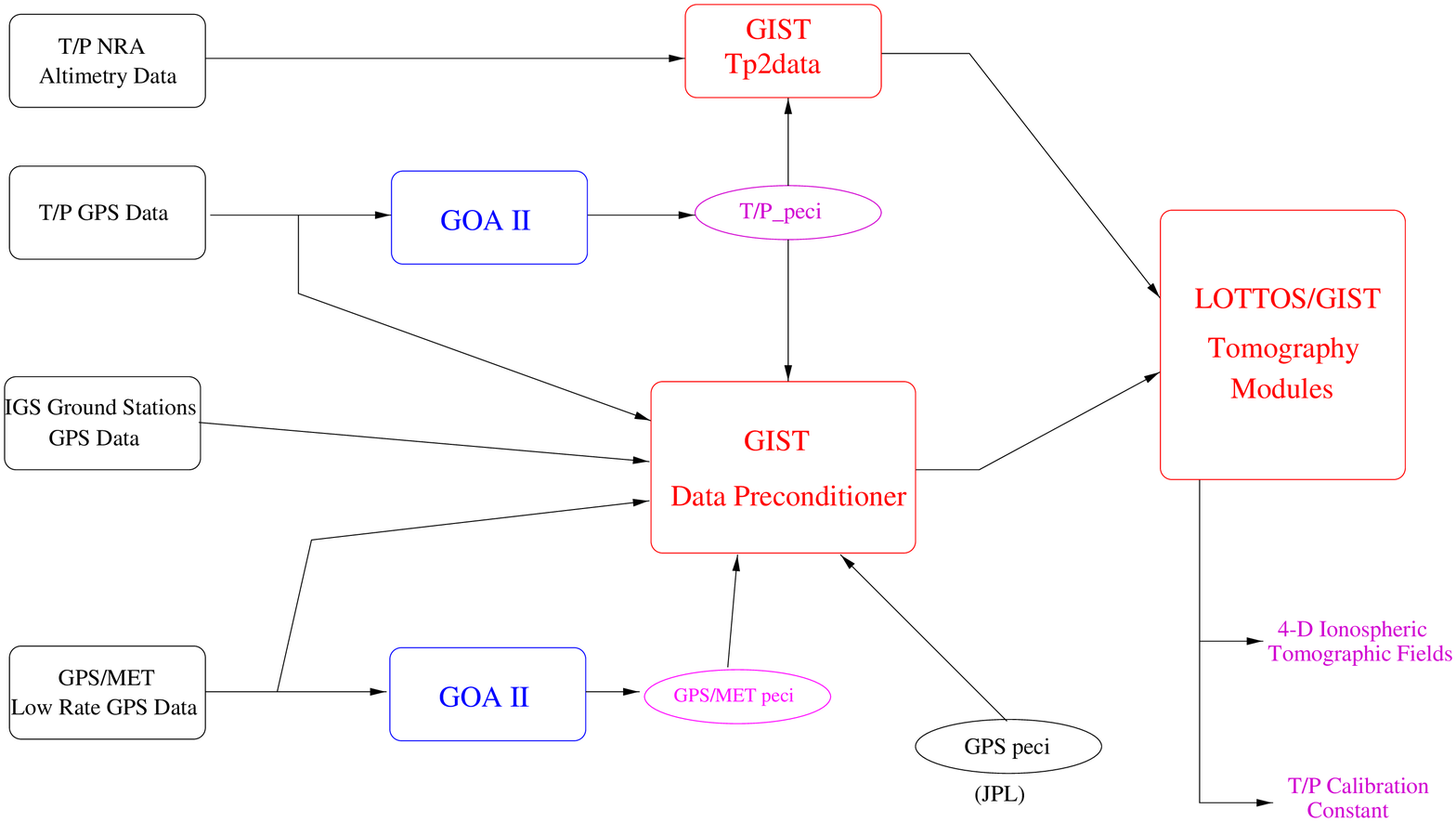}}
}
\end{center}
\caption{Block Diagram of the GIST software.}
\label{lottosgist}
\end{figure}

The software tool GIST implements the above described technique including differencing and constant estimation
strategies (for a block diagram see Figure \ref{lottosgist}). In addition, since the previous equations are valid for any dual-frequency system, different 
sources of data should be used. It has to be remembered, however, that the
tomographic solution is possible thanks to the different directions of the rays received from different
satellites which permit the system to distinguish between layers. Therefore, GPS data serve as
the basic source on which the solution is based and any additional data such as altimetric data (which is always
in the same direction) should be fed as an aiding source of information and with
the main goal of constraining the values of $\rho$ to obtain the calibration constants. In 
monostatic systems the two constants are merged into one. In this fashion, we can calibrate the instrument as part
of the overall solution.

The package GIST shares common modules with the package LOTTOS, oriented to Tropospheric Tomography (see
\cite{201flores1999}) and has the following features:
\begin{itemize}
\item Raw RINEX data conditioning: cycle slip detection, phase alignment, and data decimation.
\item Altimeter Data conditioning
\item Linear System Construction
\item Kalman Filtering with Random Walk Stochastic Process.
\item Different Constraints Strategies
\end{itemize}
The input data are GPS raw phases and pseudoranges, precise orbits for all the satellites in ECI format and
time-tagged Total Electron Contents data from other sources. In \cite{199ruffini1998} 
we discussed the convenience of the constant estimation in the data processing due to the robustness of 
the system and the existence of systematic noise sinks. However, this approach is
computationally intensive and in some cases, for system testing, it is interesting to have a rapid solution 
even if it is with low accuracy. In such cases, differencing is an attractive approach because it reduces
the number of unknowns and it is hence included as an option in the GIST package; it has to be advised, however, 
that this technique is more sensitive to systematic noise in the data or mismodeling.
\section{Results}
\begin{figure}
\begin{center}
\mbox{

\epsfxsize=10cm
\epsffile{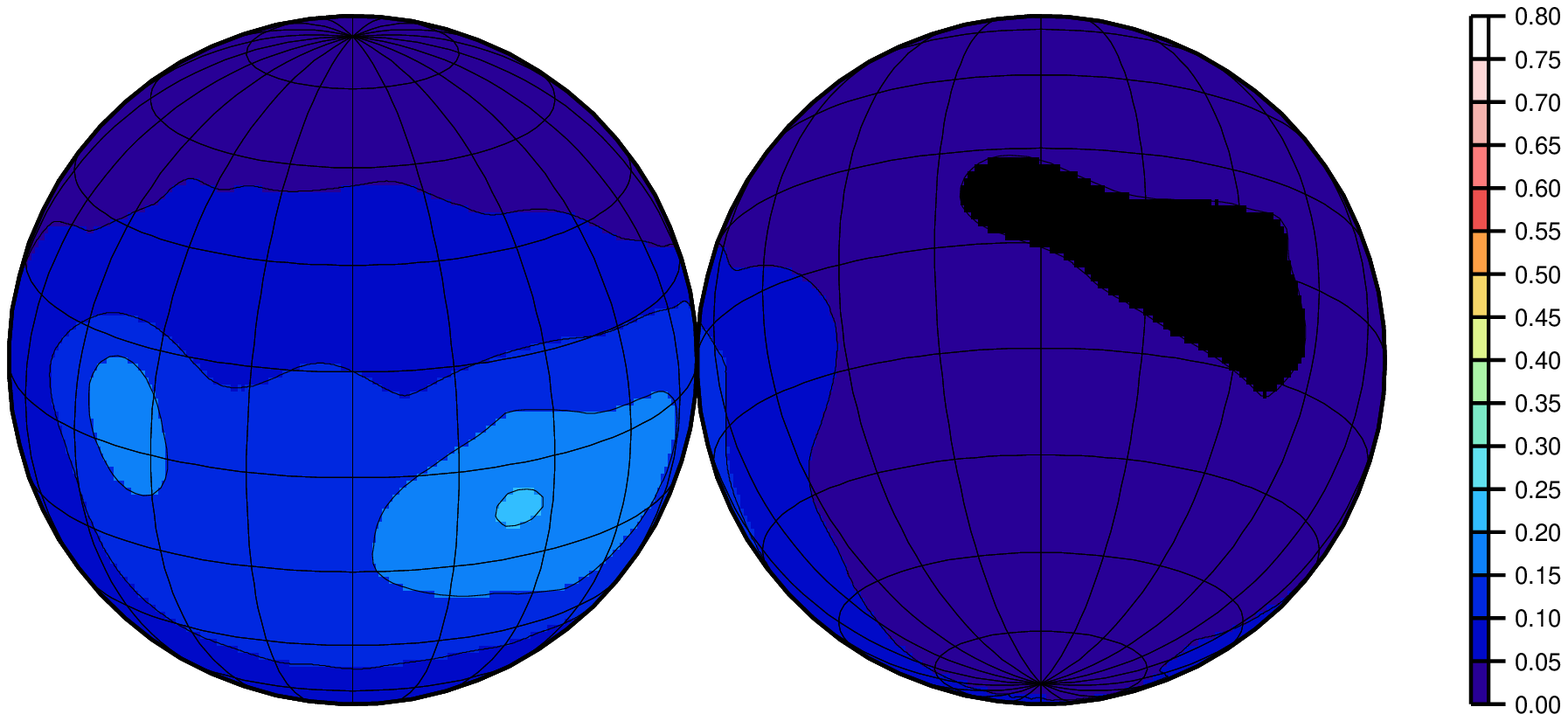}}

\mbox{
\epsfxsize=10cm
\epsffile{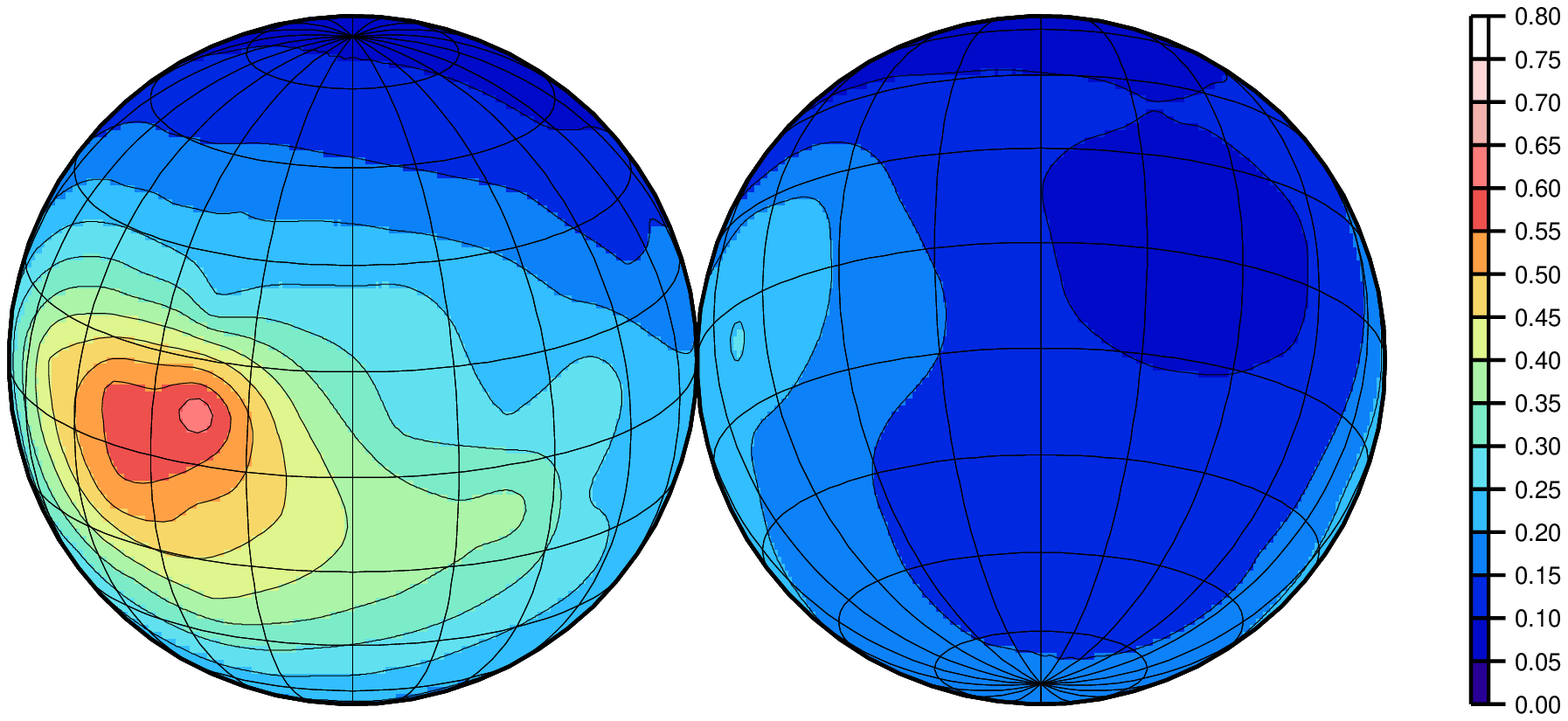}}
\mbox{
\epsfxsize=10cm
\epsffile{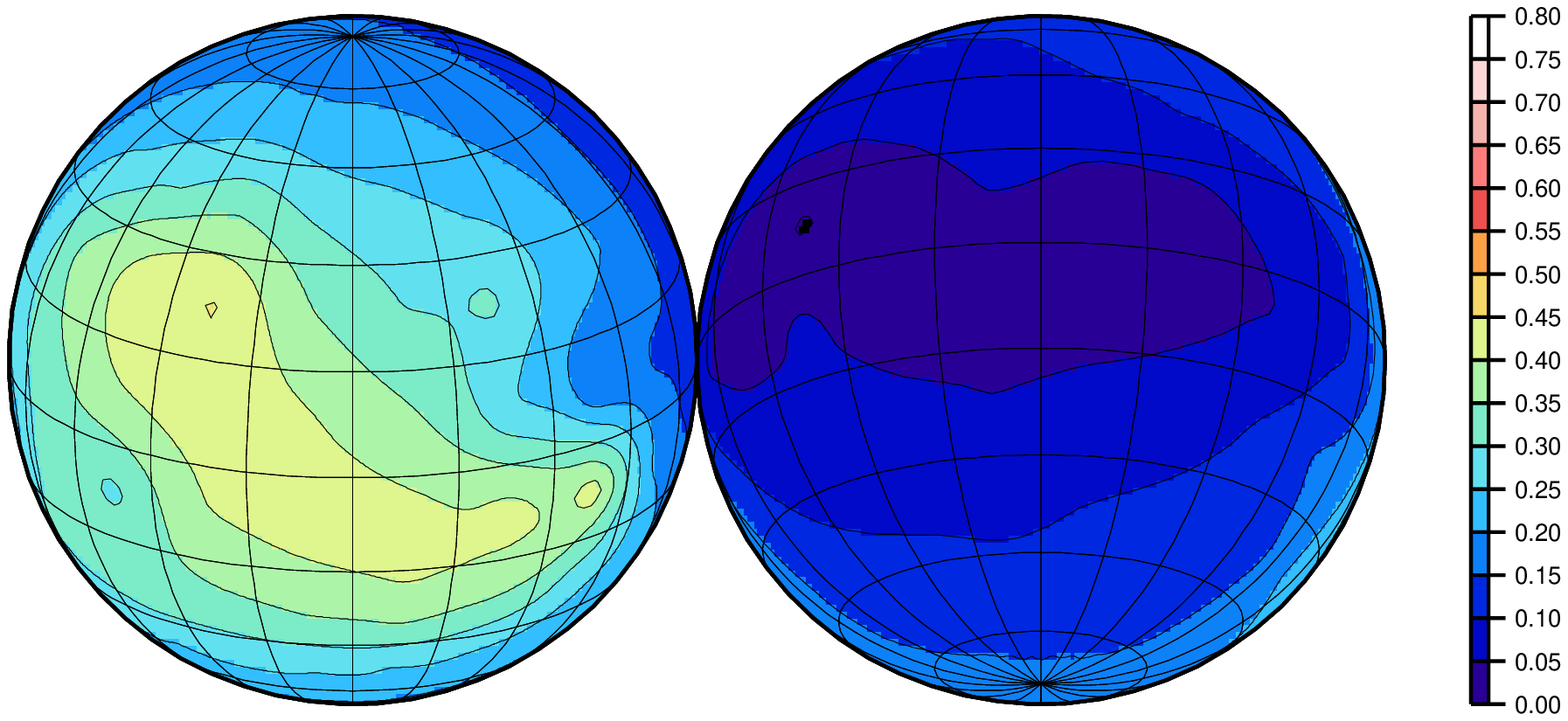}
}
\end{center}
\caption{Representation of the electron density in layers at 6575 Km, 6725 Km, and 6900 Km shells from the center of
the Earth (from bottom to top).}
\label{balls1}
\end{figure}

\begin{figure}
\begin{center}
\mbox{

\epsfxsize=10cm
\epsffile{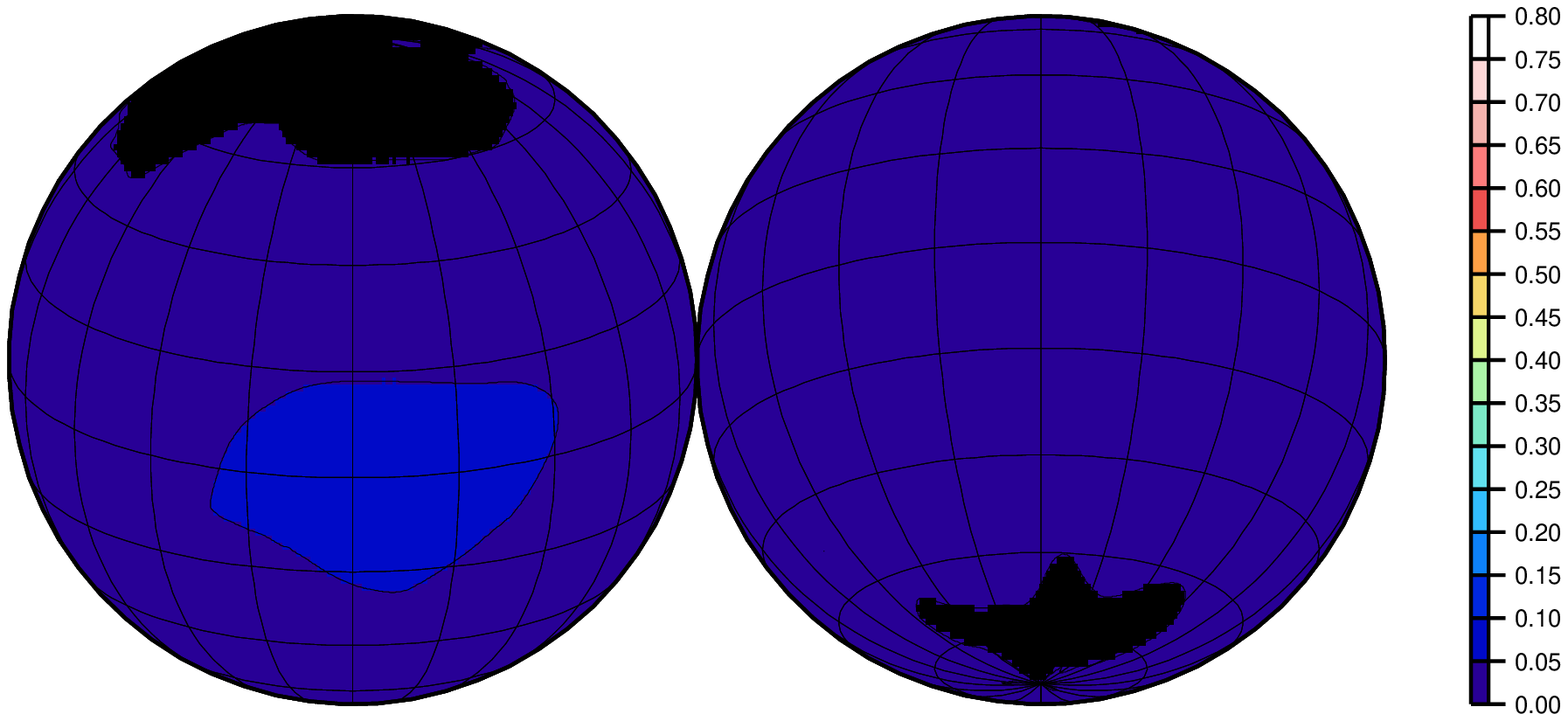}}
\mbox{

\epsfxsize=10cm
\epsffile{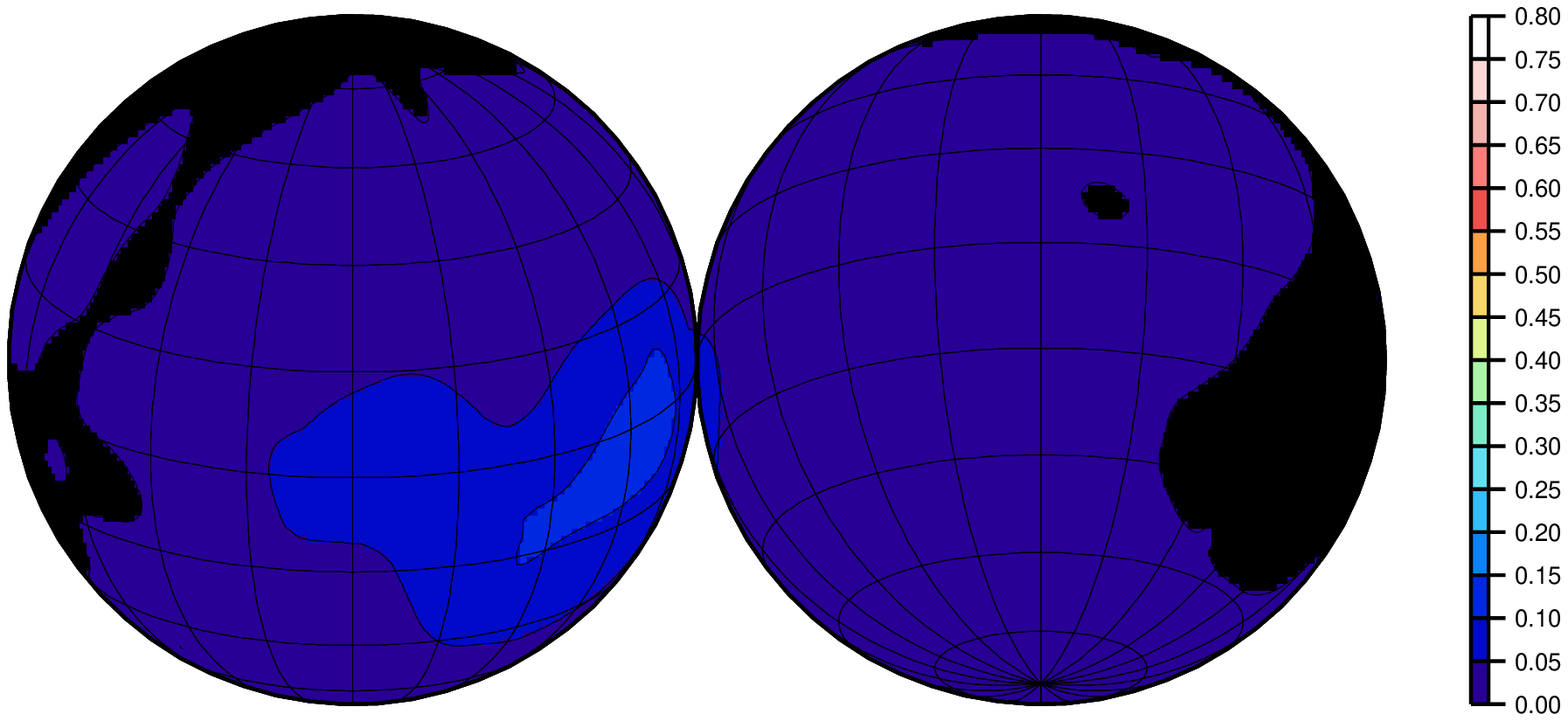}
}
\mbox{
\epsfxsize=10cm
\epsffile{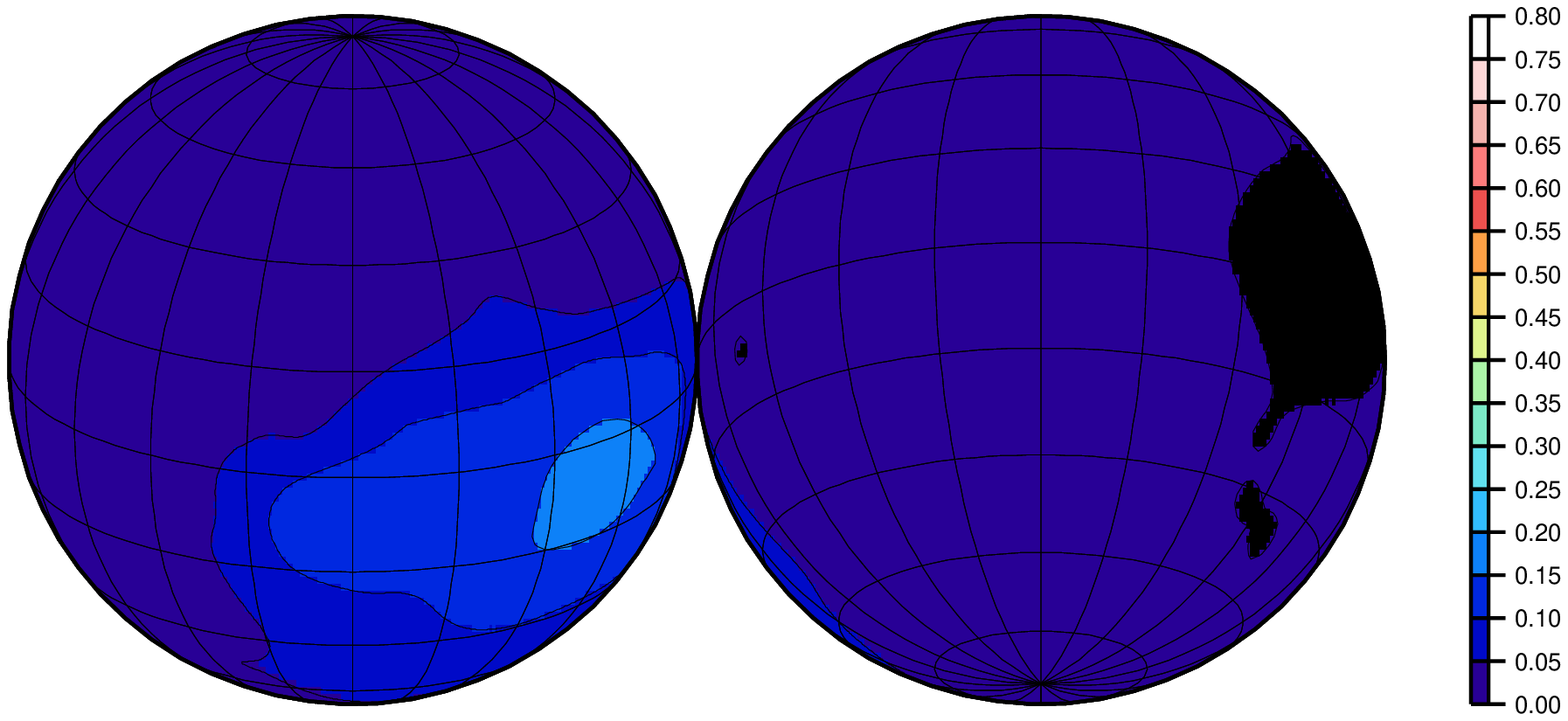}

}\end{center}
\caption{Representation of the electron density in layers at 7175 Km, 7525 Km, and 8250 Km shells from the center of
the Earth (from bottom to top).}
\label{balls2}
\end{figure}

\begin{figure}
\begin{center}
\mbox{
\epsfxsize=10cm
\epsffile{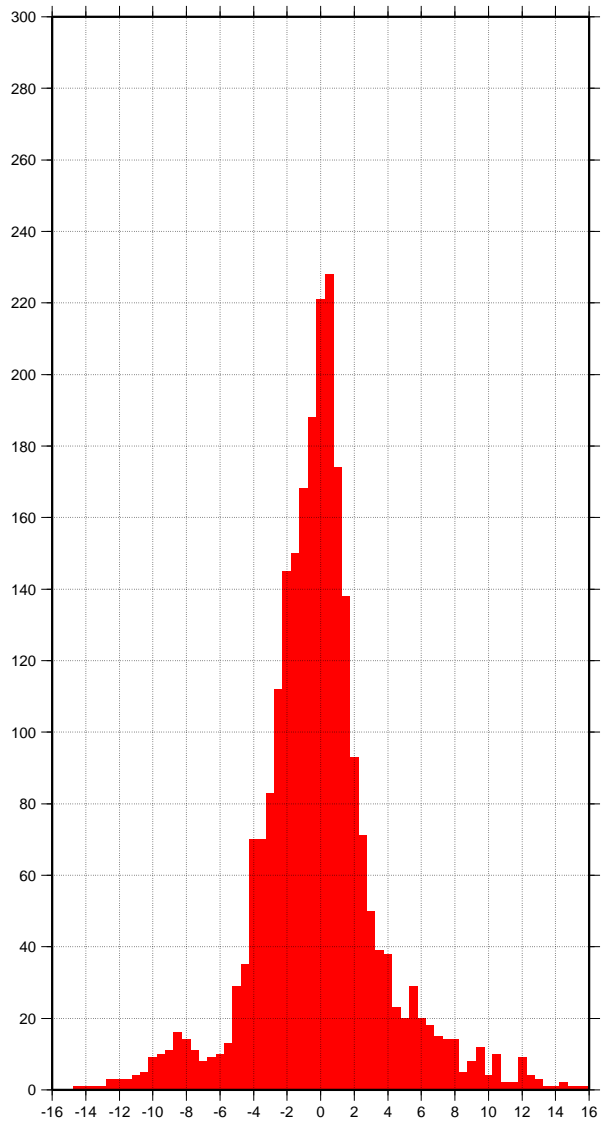}}
\end{center}
\caption{Histogram of the residues of the altimeter TEC measurements (x-axis in TECU).}
\label{residues}
\end{figure}

We have taken data from 106 IGS ground stations for 21st February 1997, GPS/MET low rate data and TOPEX/POSEIDON
data from the on-board GPS receiver (zenith-looking for navigation purposes) and the on-board NRA altimeter
data. A global grid with 20 divisions in longitude, 10 divisions in latitude and 6 layers (5 below the TOPEX/POSEIDON orbit
and 1 above to absorbe the protonosphere) has been used and the data divided into 3-hour batches for Kalman
filtering. The data were weighted according to the sigma value of the measurements (0.1 m for GPS data and 1 TECU for
TOPEX/POSEIDON \cite{75rius1997}, \cite{214imel1994}) and the orbits for the LEO were estimated using the GIPSY-OASIS II
software \cite{191webb1997}. In Figures \ref{balls1}  and \ref{balls2}  we see the 6 layers of the ionosphere, and in Figure \ref{residues} the 
residues for the T/P altimeter data. The bias constant is 2.98 TECU with a formal error of 2.58 mTECU for the T/P
Radar Altimeter, which agrees fairly well with what was reported in \cite{199ruffini1998}. 
\section{Conclusions}
We have successfully developed a solid software tool GIST for ionospheric tomography and applied it to one
day of data to yield 4D ionospheric maps. These maps are consistent with previous work and, in addition, the ingestion
of altimeter data into the model permits the direct calibration of the instrumentation. We foresee this technique
to be a very useful technique particularly when other sources of opportunity such as GPS data from satellites
or airplanes are included because of the great densification of measurements.
\section{Acknowledgements}
The authors would like to thank N. Picot (CNES), B. Haines (JPL) and C. Rocken (UCAR) for providing the data.
 This work was supported by the EC grant WAVEFRONT PL-952007 and the Comissionat per a 
Universitats i Recerca de la Generalitat de Catalunya.

\end{document}